\documentclass[pra,twocolumn,superscriptaddress]{revtex4-2}

\usepackage[utf8]{inputenc}
\usepackage[english]{babel}
\usepackage[T1]{fontenc}
\usepackage{lmodern}

\usepackage{graphicx}
\usepackage{amsmath}
\usepackage{amsfonts}
\usepackage{amssymb}
\usepackage{amsthm}

\usepackage{mathtools}

\usepackage{dsfont}
\usepackage{microtype}

\usepackage{hyperref}
\hypersetup{colorlinks=true, linkcolor=blue, citecolor=blue, urlcolor=blue}

\usepackage{ulem}
\usepackage{xcolor}

\renewcommand{\Re}{\operatorname{Re}}
\renewcommand{\Im}{\operatorname{Im}}
\newcommand{\coloneq}{\mathrel{\mathop:}=}

\newcommand{\dd}{\mathrm{d}}

\newcommand{\indexc}{\mathrm{c}}
\newcommand{\indexa}{\mathrm{a}}
\newcommand{\omegac}{\omega_\indexc}
\newcommand{\lambdac}{\lambda_\indexc}
\newcommand{\kc}{k_\indexc}
\newcommand{\Deltac}{\Delta_\indexc}
\newcommand{\Deltaa}{\Delta_\indexa}
\newcommand{\Omegac}{\Omega_\indexc}
\newcommand{\Omegap}{\Omega_\mathrm{p}}
\newcommand{\omegar}{\omega_\mathrm{r}}
\newcommand{\UR}{\hat{U}_\mathrm{R}}

\begin{document}

\title{Spin and density self-ordering in dynamic polarization gradients fields}

\author{Natalia Masalaeva}
\email{natalia.masalaeva@yandex.ru}
\affiliation{Saint Petersburg State University, 7/9 Universitetskaya nab., Saint Petersburg, 199034 Russia}
\affiliation{Institut f\"ur Theoretische Physik, Universit\"at Innsbruck, Technikerstra{\ss}e~21a, A-6020~Innsbruck, Austria}
\author{Wolfgang Niedenzu}
\affiliation{Institut f\"ur Theoretische Physik, Universit\"at Innsbruck, Technikerstra{\ss}e~21a, A-6020~Innsbruck, Austria}
\author{Farokh Mivehvar}
\affiliation{Institut f\"ur Theoretische Physik, Universit\"at Innsbruck, Technikerstra{\ss}e~21a, A-6020~Innsbruck, Austria}
\author{Helmut Ritsch}
\affiliation{Institut f\"ur Theoretische Physik, Universit\"at Innsbruck, Technikerstra{\ss}e~21a, A-6020~Innsbruck, Austria}

\begin{abstract}

We study the zero-temperature quantum phase diagram for a two-component Bose-Einstein condensate in an optical cavity. The two atomic spin states are Raman coupled by two transverse orthogonally-polarized, blue detuned plane-wave lasers inducing a repulsive cavity potential. For weak pump the lasers favor a state with homogeneous density and predefined uniform spin direction. When one pump laser is polarized parallel to the cavity mode polarization, the photons coherently scattered into the resonator induce a polarization gradient along the cavity axis, which mediates long-range density-density, spin-density, and spin-spin interactions. We show that the coupled atom-cavity system implements central aspects of the $t$-$J$-$V$-$W$ model with a rich phase diagram. At the mean-field limit we identify at least four qualitatively distinct density- and spin-ordered phases including ferro- and anti-ferromagnetic order along the cavity axis, which can be controlled via the pump strength and detuning. A real time observation of amplitude and phase of the emitted fields bears strong signatures of the realized phase and allows for real-time determination of phase transition lines. Together with measurements of the population imbalance most properties of the phase diagram can be reconstructed.

\end{abstract}

\date{\today}

\maketitle

\section{Introduction}

Quantum gas cavity QED---ultracold atoms near zero temperature coupled to photons in high-$Q$ cavity modes---has become an outstanding experimental platform to study coherent many-body quantum dynamics in a precisely controllable and readily observable form~\cite{ritsch2013cold,mekhov2012quantum}. Operating in the dispersive regime, optical atomic excitations and spontaneous emission are strongly suppressed so that coherence prevails for times long enough to observe quantum phases in great detail and study the corresponding phase transitions in real time. In essence, the non-local collective scattering of photons in and out of cavity modes by the atoms mediates long-range periodic interactions among the atoms.

\par

In a seminal experiment at the ETH Z\"urich~\cite{baumann2010dicke}, the Dicke superradiant quantum phase transition was observed almost 40 years after its prediction in the 70s~\cite{hepp1973superradiant,wang1973phase}. In a generalized setup involving an additional optical lattice, detailed measurements soon after revealed that the interplay between cavity-induced long-range density-density interactions and local contact collisional interactions leads to an even richer phase diagram including a Mott-insulator, a superfluid, a density-wave, and in particular a lattice supersolid state~\cite{landig2016quantum}.

\par

Making use of several atomic Zeeman sub-levels allows to emulate pseudospin dynamics in ultracold atomic gases. It has been suggested theoretically~\cite{gopalakrishnan2011frustration,safaei2013raman,mivehvar2017disorder, mivehvar2019cavity,ostermann2019cavity,colella2019antiferromagnetic} that cavity-enhanced Raman transitions can induce long-range periodic spin-spin interactions. Their feasibility has been experimentally confirmed soon after by several groups independently~\cite{landini2018formation,kroeze2018spinor,kroeze2019dynamical}.

\par

By tailored spatial arrangements of polarized pump lasers, a variety of long-range spin Hamiltonians can be implemented via cavity-mediated spin-spin interactions as highlighted in Ref.~\cite{mivehvar2019cavity}. Note that these cavity-induced long-range spin-spin interactions are independent of the temperature of the atomic cloud, reminiscent of dipolar interactions between polar molecules~\cite{gorshkov2011quantum,yan2013observation,wall2015quantum}, providing a promising route for simulation of quantum magnetism.

\par

Here using cavity-enhanced Raman coupling in a $\Lambda$ scheme via two external pump lasers and a cavity mode blue detuned with respect to the atomic transitions, we encounter \textit{dynamical} polarization gradients. It is known that strong local polarization gradients  in free space induce so-called non-adiabatic forces for atoms with Raman-coupled sub-levels~\cite{lacki2016nanoscale}. For our chosen parameter regimes, however, such non-adiabatic forces play only a minor role. By contrast, the polarization modulation of the effective dynamic light field along the cavity axis induces dominantly long-range interactions among the atoms via the $p$-band self-ordering~\cite{piazza2015self,zupancic2019pband,li2020measuring}. We demonstrate how these complex dynamics can be exploited to engineer \textit{combined} cavity-induced long-range ``density-density,'' ``spin-spin,'' and ``density-spin'' interactions among the effective two-component bosonic atoms~\cite{deng2014bose,mivehvar2015enhanced}. Our proposal will offer an alternative approach for simulating $t$-$J$-$V$-$W$-like models implemented via polar molecules in optical lattices~\cite{gorshkov2011tunable} with interesting topological phases~\cite{fazzini2019interaction}.

\section{Model}

\begin{figure}
  \centering
  \includegraphics[width=0.95\columnwidth]{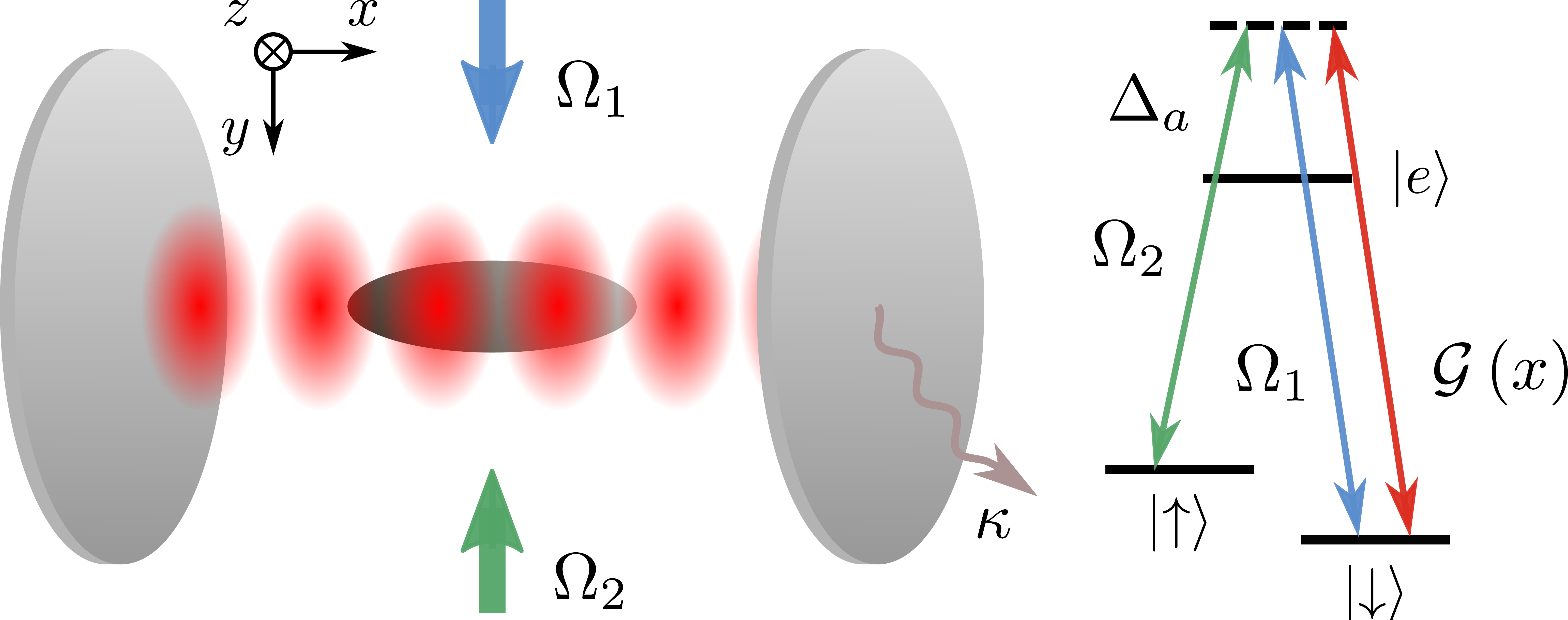}
  \caption{Schematic illustration of the transversely-pumped one-dimensional spinor BEC inside a cavity.}
  \label{scheme}
\end{figure}

Consider a cigar-shaped Bose-Einstein condensate (BEC) of $\Lambda$-type three-level atoms placed within a high-$Q$ linear cavity. The atoms are illuminated by two external pump lasers that impinge in the transverse direction as depicted in Fig.~\ref{scheme}. We assume the atomic motion to be strongly confined in the transverse directions by an additional trapping potential. The two atomic ground states $\left\{ \mid\uparrow\rangle,\mid\downarrow\rangle\right\}$ with energies $\hbar\omega_{\uparrow}$ and $\text{ \ensuremath{\hbar\omega_{\downarrow}}=\,0}$, respectively, are coupled to the excited state $|e\rangle$ with energy $\hbar\omega_{e}$ through the interaction with the cavity field and the two classical pump fields. The cavity supports one standing mode of frequency $\omegac$ with linear polarization along the $z$ axis, which is the quantization axis. This mode couples to the $\mid\downarrow\rangle\leftrightarrow|e\rangle$ transition with the position-dependent strength $\mathcal{G}(x)=\mathcal{G}_{0}\cos\left(\kc x\right)$, where $\kc =\omegac/c=2\pi/\lambdac$ is the cavity wave number. The two classical laser fields with Rabi frequencies $\Omega_{1}$ and $\Omega_{2}$ are linearly polarized along the $z$ and $x$ axis,  respectively, driving the transitions $\mid\downarrow\rangle\leftrightarrow|e\rangle$ and $\mid\uparrow\rangle\leftrightarrow|e\rangle$ (Fig.~\ref{scheme}). The total electric field
\begin{multline}\label{E_tot}
  \hat{\mathbf{E}}(x,y)=E_1 e^{ik_1y}\mathbf{e}_z + E_2 e^{-ik_2y}\mathbf{e}_x \\
  +\sqrt{\frac{\hbar\omegac}{2\epsilon_{0} V}}\hat{a}\cos\left(\kc x\right)\mathbf{e}_z + \mathrm{H.c.}
\end{multline}
thus features polarization gradient along the cavity axis as the orientation of the polarization vector depends on the position $x$. Here $E_1$ and $E_2$ are complex amplitudes of the classical fields, $\hat{a}$ is the bosonic annihilation operator for the  photonic field, $V$ the cavity quantization volume, and $\epsilon_0$ the vacuum permittivity.

\par

The cavity and laser frequencies $\{\omegac, \omega_{1},\omega_{2}\}$ are assumed to be blue detuned with respect to the atomic transition frequencies, e.g., $\omega_{1}-\omega_{e}\gg0$, while pump frequencies are close to resonant with one another, e.g., $|\omega_{1}-\omega_{2}|/\omega_{1(2)}\ll1$. This implies that the atoms are attracted to the intensity minima of the light fields~\cite{domokos2003mechanical} and would, in general, lead to the suppression of light scattering into the resonator. As we show in this work, for some pump strengths, however, a self-ordered phase is still generated due to the complex interplay of collective coherent scattering and optical dipole forces. Similar features were recently also found for spinless BECs via the $p$-band coherent photon scattering~\cite{piazza2015self,zupancic2019pband,li2020measuring} and polarisable point particles~\cite{jungkind2019optomechanical}.

\par

In typical cavity-QED experiments, the condensates are rather dilute so that local collisional contact interactions are negligible compared to the cavity-mediated long-range interactions. For that reason we do not include two-body contact interactions in our model~\cite{leonard2017supersolid, schuster2018pinning}.

\par

\begin{figure*}
  \centering
  \includegraphics[width=0.95\textwidth]{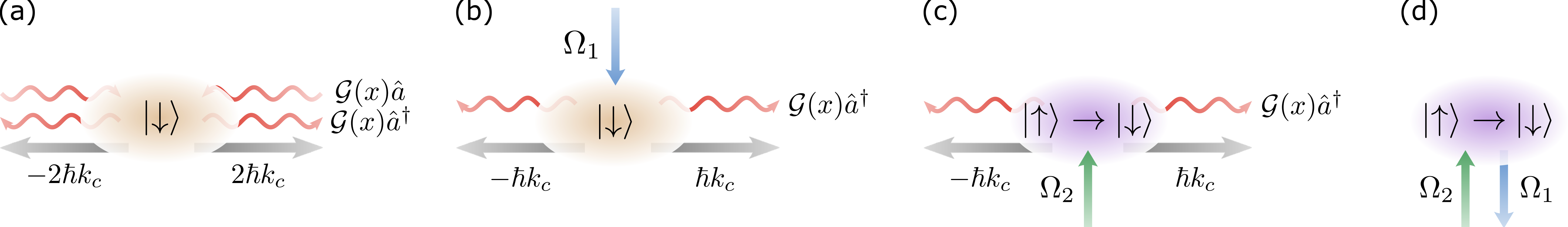}
  \caption{Schematic representation of the two-photon processes that contribute to the Hamiltonian density~\eqref{1particle_H}. The scattering events devoid of a spin flip that are comprised in Eq.~\eqref{eq_Udown} consist of (a)~the absorption and re-emission of cavity photons and (b)~the scattering of a photon from the pump laser $\Omega_1$ into the cavity mode. Atomic pseudospin flips $\mid\uparrow\rangle\mapsto\:\mid\downarrow\rangle$ as described in Eq.~\eqref{eq_UR} occur in photon scattering events (c)~from the pump laser $\Omega_2$ into the cavity mode or (d)~from the first into the second pump laser. All processes that involve the cavity field [(a)--(c)] cause a change in the atomic momentum distribution (grey arrows). In the one-dimensional description we assume that any transverse momenta are absorbed by an external trapping potential.}
  \label{fig_2_processes}
\end{figure*}

\par

In the limit of large atom-pump detuning, the atomic excited state can be adiabatically eliminated~\cite{ritsch2013cold}, leading to the effective many-body Hamiltonian (see Appendix~\ref{A})
\begin{equation}\label{eq_H}
  \hat{H}=\int\hat{\Psi}^{\dagger}(x)\tilde{\mathcal{H}}\hat{\Psi}(x)\dd x-\hbar\Deltac\hat{a}^{\dagger}\hat{a},
\end{equation}
where $\hat{\Psi}\coloneq(\hat{\psi}_{\uparrow},\hat{\psi}_{\downarrow})^{\top}$ is the bosonic annihilation operators for the spinor atomic fields, and $\Deltac\coloneq\omega_{1}-\omegac<0$ is the cavity detuning. As the detuning $\Deltac$ is negative, the atomic kinetic energy decreases in a single scattering event, which leads to atomic cooling of fast atoms~\cite{ritsch2013cold}. In the $\{\uparrow,\downarrow\}$ basis, the single-particle Hamiltonian density has the matrix representation
\begin{equation}
  \tilde{\mathcal{H}}=\left(\begin{array}{cc}
-\frac{\hbar^2}{2m}\partial^2_x+\hbar\tilde{\delta} & \hbar\UR(x)\\
\hbar\UR^{\dagger}(x) & -\frac{\hbar^2}{2m}\partial^2_x+\hbar\hat{U}_{\downarrow}(x)
\end{array}\right),\label{1particle_H}
\end{equation}
where $\tilde{\delta}\coloneq\omega_{\uparrow}-\left(\omega_{1}-\omega_{2}\right)+\Omega_{2}^{2}/\Deltaa-\Omega_{1}^{2}/\Deltaa$ is the Stark-shifted two-photon detuning with $\Deltaa\coloneq\omega_{1}-\omega_{e}$ and the operators $\hat{U}_{\downarrow}(x)$ and $\UR^{\left(\dagger\right)}(x)$ describe various atom-light interactions (see Fig~\ref{fig_2_processes}):

(i)~The scattering of photons by the spin-$\left|\downarrow\right\rangle$ atoms without changing their internal state results in the $\lambdac$-periodic dynamical potential
\begin{equation}\label{eq_Udown}
  \hbar\hat{U}_{\downarrow}(x)= \hbar U_{0}\hat{a}^{\dagger}\hat{a}\cos^{2}\left(\kc x\right)+\hbar \eta\left(\hat{a}+\hat{a}^{\dagger}\right)\cos\left(\kc x\right).
\end{equation}
Its first contribution accounts for the absorption and re-emission of cavity photons by the atoms, see Fig \ref{fig_2_processes}(a), where $U_{0}\coloneq\mathcal{G}_{0}^{2}/\Deltaa$ is the maximum depth of this potential per photon. The second contribution, depicted in Fig.~\ref{fig_2_processes}(b), describes the coherent scattering of photons between the transverse pump $\Omega_{1}$ and the cavity mode with effective strength $\eta\coloneq\mathcal{G}_{0}\Omega_{1}/\Deltaa$.

\par

(ii)~The processes that are accompanied by an atomic pseudospin flip induce the $\lambdac$-periodic, dynamical two-photon Raman coupling
\begin{equation}\label{eq_UR}
  \hbar\UR^{\left(\dagger\right)}(x)=\hbar\Omegac\hat{a}^{\left(\dagger\right)}\cos\left(\kc x\right)+\hbar\Omegap,
\end{equation}
that describes the exchange of photons between the second pump laser and the cavity field [Fig.~\ref{fig_2_processes}(c)] and between the two pump lasers [Fig.~\ref{fig_2_processes}(d)], respectively. These scattering events occur with effective Raman coupling strengths
\begin{equation}\label{eq_Omegacp}
  \Omegac\coloneq\mathcal{G}_{0}\Omega_{2}/\Deltaa, \qquad \Omegap\coloneq\Omega_{1}\Omega_{2}/\Deltaa.
\end{equation}
Without loss of generality, we have assumed $\{\mathcal{G}_0,\Omega_1,\Omega_2\}\in\mathbb{R}_+$.

\par

The single-particle Hamiltonian density~\eqref{1particle_H} possesses a discrete $\mathbb{Z}_2$ symmetry. Namely, it is invariant under a simultaneous spatial translation $x\mapsto x+\lambdac/2$ and a parity transformation of the field amplitude $\hat a \mapsto -\hat a$. This is the same symmetry as for transversally-pumped two-level (i.e., effectively single-component) atoms in linear resonators. There, this symmetry is spontaneously broken above a certain critical pump strength, which is known as self-organisation~\cite{domokos2002collective,asboth2005self}. Hence, we expect a similar symmetry breaking in our spinor BEC system. Owing to its more complex level structure, however, the intricate interplay between the atomic density, the atomic pseudospin, and the cavity mode leads to richer phase diagrams than for spinless particles~\cite{nagy2008self,baumann2010dicke,niedenzu2011kinetic,piazza2013bose,sandner2015selfordered,schuetz2015thermodynamics,landig2016quantum,zupancic2019pband}, as we are going to discuss in the following.

\section{Cavity induced long-range interactions and the Effective spin Hamiltonian}

For large cavity detuning $|\Deltac|$ and/or large photon decay rate $2\kappa$ the light field instantly follows the atomic distribution and quickly attains its  steady state~\cite{nagy2008self}
\begin{equation}
  \hat{a}_\mathrm{ss}=\frac{\eta\int\cos\left(\kc x\right)\hat{n}_{\downarrow}\dd x+\Omegac\int\cos\left(\kc x\right)\hat{s}_{-}(x)\dd x}{\Deltac+i\kappa-U_{0}\int\cos^{2}\left(\kc x\right)\hat{n}_{\downarrow}\dd x},\label{steady_state}
\end{equation}
where $\hat{n}_{\tau}(x)=\hat{\psi}_{\tau}^{\dagger}(x)\hat{\psi}_{\tau}(x)$ is the local atomic density operator of state $\tau\in\{\uparrow,\downarrow\}$ and $\hat{s}_{-}(x)=\hat{\psi}_{\downarrow}^{\dagger}(x)\hat{\psi}_{\uparrow}(x)$ is the local atomic spin lowering operator. The cavity field is hence coupled to the atomic density $\hat{n}_{\downarrow}$ and the atomic spin polarization $\hat{s}_{-}(x)$, in contrast to the conventional self-ordering of single component BECs~\cite{nagy2008self} or spinor BECs~\cite{mivehvar2017disorder}. The emission of photons into the cavity field mode by the atoms may either leave the atomic internal state untouched [first term in the numerator of Eq.~\eqref{steady_state}] or induce a pseudospin flip $\mid\uparrow\rangle\mapsto\:\mid\downarrow\rangle$ [second term in the numerator of Eq.~\eqref{steady_state}].

\par

Substituting the steady-state light field~\eqref{steady_state} into the many-body Hamiltonian~\eqref{eq_H} leads to an effective spin Hamiltonian
\begin{equation}\label{H_spin}
  \hat{H}_{\mathrm{spin}}=\hat{H}_\mathrm{kin}+\hat{H}_{J\text{-}V\text{-}W}+\hat{H}_{xz},
\end{equation}

where $\hat{H}_\mathrm{kin}$ is the kinetic energy, and

\begin{widetext}
  \begin{subequations}\label{eq_HJVW_Hxz}
  \begin{align}\label{eq_H_JVW}
 \hat{H}_{J\text{-}V\text{-}W}&=\iint\Big\{ J_{\bot}\left(x,x'\right)\big[\hat{s}_{x}(x)\hat{s}_{x}\left(x'\right)+\hat{s}_{y}(x)\hat{s}_{y}\left(x'\right)\big]+J_{z}\left(x,x'\right)\hat{s}_{z}(x)\hat{s}_{z}\left(x'\right)\Big\} \dd x\dd x'+\int\mathbf{B}\cdot\hat{\mathbf{s}}(x)\dd x\notag\\ &\quad+\iint V\left(x,x'\right)\hat{n}(x)\hat{n}\left(x'\right)\dd x\dd x'\notag\\ &\quad+\iint\Big\{ W_{x}\left(x,x'\right)\big[\hat{n}(x)\hat{s}_{x}\left(x'\right)+\hat{s}_{x}\left(x'\right)\hat{n}(x)\big]-W_{z}\left(x,x'\right)\big[\hat{n}(x)\hat{s}_{z}\left(x'\right)+\hat{s}_{z}\left(x'\right)\hat{n}(x)\big]
\Big\}\dd x\dd x',
  \end{align}
  and
  \begin{equation}\label{eq_Hxz}
    \hat{H}_{xz}=-\iint J_{xz}\left(x,x'\right)\big[\hat{s}_{z}(x)\hat{s}_{x}\left(x'\right)+\hat{s}_{x}\left(x'\right)\hat{s}_{z}(x)\big]\dd x\dd x'.
  \end{equation}
\end{subequations}
\end{widetext}
Here we have introduced the total local density operator $\hat{n}(x)=\hat{n}_{\uparrow}(x)+\hat{n}_{\downarrow}(x)$ and the local pseudospin operator $\hat{\mathbf{s}}(x)=\left(\hat{s}_{x}(x),\hat{s}_{y}(x),\hat{s}_{z}(x)\right)^\top=\hat{\Psi}^{\dagger}(x)\boldsymbol{\sigma}\hat{\Psi}(x)$, where $\boldsymbol{\sigma}=\left(\sigma_{x},\sigma_{y},\sigma_{z}\right)^\top$ are the Pauli matrices.

\par

The Hamiltonian $ \hat{H}_{J\text{-}V\text{-}W}$, Eq.~\eqref{eq_H_JVW}, together with the kinetic energy $\hat{H}_\mathrm{kin}$ corresponds to a long-range anisotropic \textit{t-J-V-W} model \cite{gorshkov2011tunable}. The first line of Eq.~\eqref{eq_H_JVW} corresponds to a long-range XXZ Heisenberg spin Hamiltonian with an effective homogeneous magnetic field
\begin{equation}\label{eq_B}
  \mathbf{B=\hbar\mathrm{\left(\frac{2\Omega_{1}\Omega_{2}}{\Deltaa},0,\tilde{\delta}\right)}}^\top.
\end{equation}
Its $x$ component thereby originates from the Raman coupling between the two pump lasers and its $z$ component stems from the effective detuning between the two pseudospin states. The second and third lines of $\hat{H}_{J\text{-}V\text{-}W}$ contain long-range density-density and density-spin interactions, respectively. Finally, the Hamiltonian $\hat{H}_{xz}$, Eq.~\eqref{eq_Hxz}, describes the long-range cross-couplings between $x$ and $z$ spin components.

\par

We note that all coupling coefficients in Eqs.~\eqref{eq_HJVW_Hxz} share the same position dependence
\begin{equation}
  c\left(x,x'\right)=\frac{\hbar\big[2\Re\hat{\tilde{\Delta}}_\mathrm{c}-\Deltac\big]\mathcal{G}_{0}^{2}}
  {2|\hat{\tilde{\Delta}}_\mathrm{c}|^{2}\Deltaa^{2}}
  \cos\left(\kc x\right)\cos\left(\kc x'\right),
\end{equation}
with $\hat{\tilde{\Delta}}_\mathrm{c}=\Deltac+i\kappa-U_{0}\int\cos^{2}\left(\kc x\right)\hat{n}_{\downarrow}\dd x$; namely,
  \begin{align}\label{coup_coeff}
    J_{\bot}\left(x,x'\right)&=2\Omega_{2}^{2}c\left(x,x'\right), \nonumber\\
    J_{z}\left(x,x'\right)&=4V\left(x,x'\right)=2W_{z}\left(x,x'\right)=2\Omega_{1}^{2}c\left(x,x'\right),\nonumber\\
    J_{xz}\left(x,x'\right)&=2W_{x}\left(x,x'\right)=2\Omega_{1}\Omega_{2}c\left(x,x'\right).
\end{align}

\par

The long-range cavity-induced interactions between the atoms thus transform the BEC to an array of itinerant interacting spins
governed by the Hamiltonian~\eqref{H_spin} that, depending on the parameters, implements different spin models, with all spin-coupling coefficients, Eq.~\eqref{coup_coeff}, widely tunable through the Rabi frequencies $\Omega_{1,2}$ of the pump lasers. Choosing $J_{z}=V=W_{x,z}=J_{xz}=0$, one can observe enhanced superconductivity~\cite{manmana2017correlations} and $d$-wave superfluidity~\cite{kuns2011dwave}. Our model system therefore opens the possibility to study various quantum magnetic phases whose magnetic order can be detected through the cavity field emission in real time.

\par

For nonzero average field in the cavity, different cavity-induced interactions in Eqs.~\eqref{eq_HJVW_Hxz} compete with one another. Depending on operation parameters, a specific interaction can be made dominant and determine the system behavior. In particular the sign of the coupling coefficients $J_{\bot}$ and $J_{z}$ sets the magnetic ordering of the spins to either a ferromagnetic (FM) or antiferromagnetic (AFM) pattern. In the next section we will characterize the expected spin textures in various limiting cases in more detail.

\section{Mean-field phase diagram}

In the mean-field regime the quantum fluctuations are omitted and the atomic and cavity field operators are replaced by their corresponding quantum averages, $\hat{\psi}_{\tau}(x,t)\rightarrow\langle\hat{\psi}_{\tau}\left(x,t\right)\rangle\equiv\psi_{\tau}\left(x,t\right)=\sqrt{n_{\tau}\left(x,t\right)}e^{i\phi_{\tau}\left(x,t\right)}$ and $\hat{a}\left(t\right)\rightarrow\langle\hat{a}\left(t\right)\rangle\equiv\alpha\left(t\right)=|\alpha(t)|e^{i\phi_{\alpha}\left(t\right)}$. The system is then described by three coupled nonlinear equations
\begin{subequations}\label{system_of_eq}
\begin{align}
  i\frac{\partial}{\partial t}\alpha&=-\tilde{\Delta}_\mathrm{c}\alpha
 +\eta\Theta +\Omegac\Xi,\\
  i\hbar\frac{\partial}{\partial t}\psi_{\uparrow}&=
  \left[-\frac{\hbar^2}{2m}\partial^2_x+\hbar\tilde{\delta}\right]\psi_{\uparrow}
  +\hbar U_{\rm R}(x)\psi_{\downarrow}, \nonumber\\
  i\hbar\frac{\partial}{\partial t}\psi_{\downarrow}&=
  \left[-\frac{\hbar^2}{2m}\partial^2_x+\hbar U_{\downarrow}(x)\right]\psi_{\downarrow}
  +\hbar U_{\rm R}^*(x)\psi_{\uparrow} \label{eq:coupled-GPE}.
\end{align}
\end{subequations}
where $U_{\downarrow}(x,\alpha)=U_{\downarrow}(x)=\langle \hat{U}_{\downarrow}(x) \rangle$ and $U_{\rm R}(x,\alpha)=U_{\rm R}(x)=\langle \UR (x) \rangle$ are the quantum averages of the corresponding operators in Eqs.~\eqref{eq_Udown} and \eqref{eq_UR}, respectively, and $\tilde{\Delta}_\mathrm{c}=\langle \hat{\tilde{\Delta}}_\mathrm{c} \rangle$. Here we have introduced the mean-field density order parameter
  \begin{equation}\label{eq_theta}
    \Theta \coloneq \int n_{\downarrow}(x)\cos\left(\kc x\right)\dd x,
  \end{equation}
  that describes the $\lambdac$-periodic spatial modulation of the (spin-$\downarrow$) atoms, and the mean-field spin order parameter
  \begin{align}\label{eq_xi}
    \Xi &\coloneq \int s_-(x)\cos\left(\kc x\right) \dd x\nonumber\\
    &=\int \left[s_x(x)-is_y(x)\right]\cos\left(\kc x\right) \dd x.
  \end{align}
Note also that only the total number of the atoms is conserved, i.e., $\sum_{\tau}\int n_{\tau}(x)\dd x=N$.

\par

The total energy of the system can be obtained as $E=-\hbar\Deltac|\alpha|^{2}+\int\mathcal{E}(x)\dd x$~\cite{mivehvar2017disorder}, where $\mathcal{E}(x)$ is the energy-functional density,
\begin{multline}
\mathcal{E}(x)=\frac{\hbar^{2}}{2m}\left(\psi_{\uparrow}^{*}\partial^2_{x}\psi_{\uparrow}+\psi_{\downarrow}^{*}\partial^2_{x}\psi_{\downarrow}\right)+\hbar\tilde{\delta}n_{\uparrow}\\
+\Big[\hbar U_{0}|\alpha|^{2}\cos^{2}\left(\kc x\right)+2\hbar\eta|\alpha|\cos\phi_{\alpha}\cos\left(\kc x\right)\Big]n_{\downarrow}\\
+2\sqrt{n_{\uparrow}n_{\downarrow}}\Big[\hbar\Omegac|\alpha|\cos\left(\phi_{\alpha}+\Delta\phi\right)\cos\left(\kc x\right)+\hbar\Omegap\cos\Delta\phi\Big],\label{energy-functional}
\end{multline}

with $\Delta\phi\coloneq\phi_{\downarrow}-\phi_{\uparrow}$ being the relative phase of the two condensate wave functions. Note that since $\Deltaa>0$, therefore $\{U_{0},\eta,\Omegac,\Omegap\}\geqslant0$.

\par

Below threshold where $\alpha=0$ the Raman coupling energy $\propto\Omegap\cos\Delta\phi$ fixes the relative phase of the spatially-homogeneous BEC to $\Delta\phi=\pi$ (recall $\Omegap>0$). Namely, the energy is minimized if two spin states have opposite phase. This is in contrast to the system studied in ref.~\cite{mivehvar2017disorder} where the relative phase could be chosen freely.

\par

\begin{figure*}
  \centering
  \includegraphics[width=0.95\textwidth]{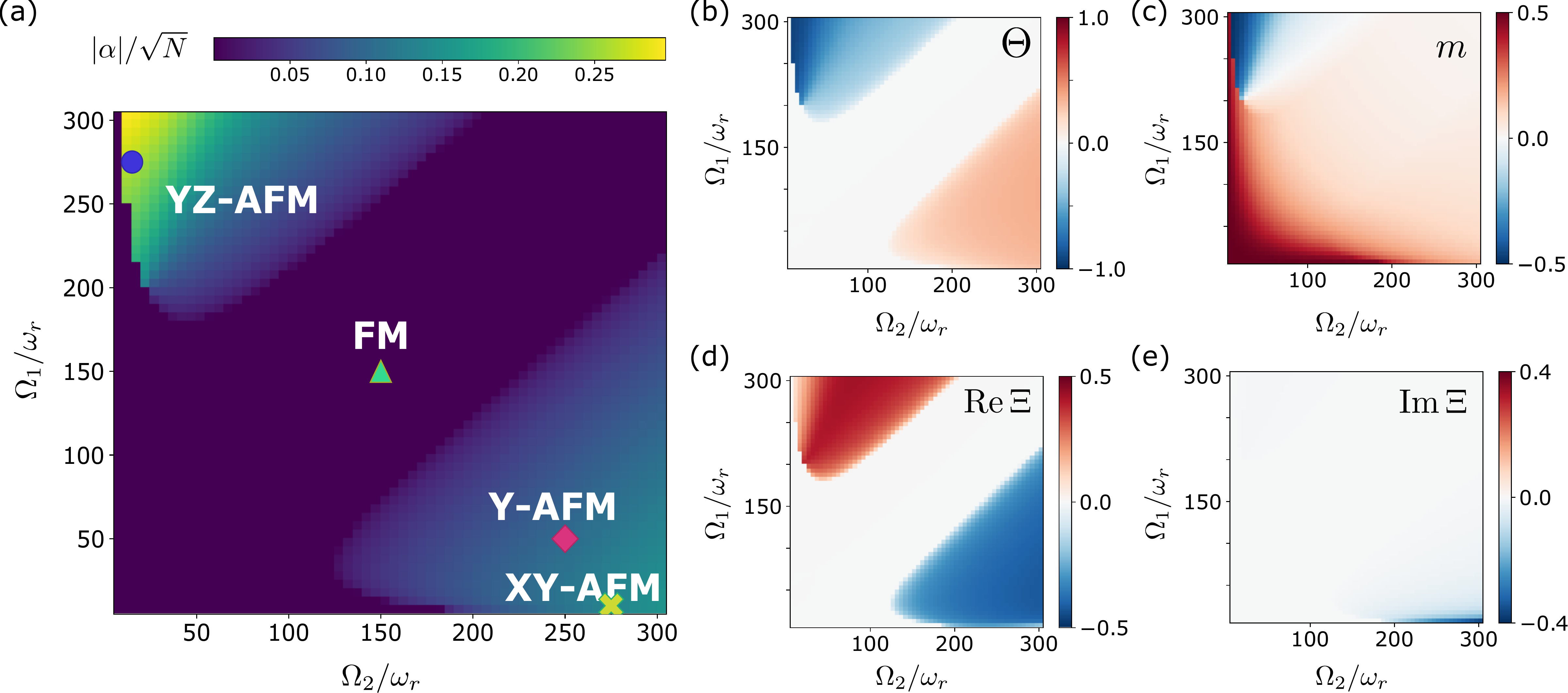}
  \caption{(a)~Mean-field phase diagram as a function of the two pump Rabi frequencies $\Omega_{1}/\omegar$ and $\Omega_{2}/\omegar$. It reveals four distinct phases: YZ-antiferromagnetic (YZ-AFM), ferromagnetic (FM), Y-antiferromagnetic (Y-AFM), and XY-antiferromagnetic (XY-AFM). The color code depicts the rescaled cavity field amplitude $|\alpha|/\sqrt{N}$. (b)--(e)~Order parameters and global magnetization: (b)~density order parameter $\Theta$ [Eq.~\eqref{eq_theta}], (c)~global magnetization $m$ [Eq.~\eqref{eq_magnetization}], (d)~real and (e)~imaginary parts of the spin order parameter $\Xi$ [Eq.~\eqref{eq_xi}], respectively. The other parameters are $\left(\Deltaa,\Deltac,\,NU_{0},\,\tilde{\delta},\,\kappa\right)=\left(10^3,\,-150,\,40,\,-5,\,25\right)\omegar$.}
  \label{phase_diagram}
\end{figure*}

Above threshold, however, the atoms scatter photons from the pump lasers into the cavity mode such that $\alpha\neq 0$. The minimization of the spatial-dependent Raman coupling energy [third line in Eq.~\eqref{energy-functional}] then results in a position-dependent relative condensate phase $\Delta\phi(x)$. As $\cos\left(\kc x\right)$ changes sign depending on the atomic position, the relative phase smoothly varies in space. Such spatial dependence of the relative condensate phase for colored markers in the phase diagram in Fig.~\ref{phase_diagram}(a) is depicted in the insets of Fig.~\ref{spin_textures}(a)--(d) and leads to intriguing phenomena in the spin structure, which will be shown below.

\par

\par

In order to obtain the mean-field phase diagram we self-consistently compute the stationary state of the cavity-field amplitude $\partial\alpha/\partial t=0$ [cf.\ Eq.~\eqref{steady_state}],
 \begin{equation}
  \alpha=\frac{1}{\tilde{\Delta}_{\rm c}}(\eta\Theta +\Omegac\Xi),\label{ss_with_order_parameters}
\end{equation}
and the corresponding atomic ground state from the coupled Schr\"odinger equations, Eq.~\eqref{eq:coupled-GPE}. Note that the coupled Schr\"odinger equations depend parametrically on $\alpha$ via $U_{\downarrow}(x)=U_{\downarrow}(x,\alpha)$ and $U_{\rm R}(x)=U_{\rm R}(x,\alpha)$, indicating the highly nonlinear nature of the system.

\par

Before proceeding to the main results of our work, let us brieﬂy clarify the issue regarding the temperature of the gas at steady state. Since our system is driven-dissipative, the stationary temperature would be non-zero and limited by the cavity decay rate. The cavity-induced atomic redistribution can even lead to non-thermal steady states. However, due to the long-range cavity-mediated interactions, the corrections to the noiseless mean-ﬁeld approach are suppressed by a factor $1/V$~\cite{piazza2013bec}, with $V$ being the volume of the atomic cloud. Then the characteristic time for the cavity-induced atomic redistribution scales with the volume $V$. Therefore, in the thermodynamic limit $N,V\rightarrow\infty$ with $N/V=\text{const}$, this characteristic time exceeds a typical experimental time scale~\cite{piazza2014quantum}, and the mean-ﬁeld description becomes exact.

\par

Figure~\ref{phase_diagram}(a) shows the mean-field cavity-field amplitude $|\alpha|/\sqrt{N}$ as a function of the two pump Rabi frequencies $\Omega_{1}/\omegar$ and $\Omega_{2}/\omegar$, where $\omegar\coloneq\hbar \kc ^{2}/2m$ is the recoil frequency. The nonzero density order parameter $\Theta$, Eq.~\eqref{eq_theta}, shown in Fig.~\ref{phase_diagram}(b) reveals the $\lambdac$-periodic atomic self-organization. The sign of $\Theta$ reflects the localization of the atoms either on even ($\Theta>0,~\kc x = 2\pi \ell$ with $\ell\in\mathbb{Z}$) or odd [$\Theta<0,~\kc x = \pi (2\ell+1)$] sites. The global magnetization of the atomic gas~\cite{colella2019antiferromagnetic}
\begin{equation}
  m \coloneq \frac{N_{\uparrow}-N_{\downarrow}}{N},
  \label{eq_magnetization}
\end{equation}
with $N_\tau \coloneq \int n_\tau (x)\dd x $ is depicted in Fig.~\ref{phase_diagram}(c). As the Stark-shifted detuning between the ground levels is chosen as $\tilde{\delta}=-5\omegar$, atoms mostly prefer to occupy the $\mid\uparrow\rangle$ ground state, except a region where $\Omega_1$ is much larger than $\Omega_2$. The real and imaginary parts of the spin order parameter $\Xi$ are illustrated in Figs.~\ref{phase_diagram}(d) and~\ref{phase_diagram}(e), respectively. Noting Eq.~\eqref{eq_xi}, a non-zero $\Re\Xi$ ($\Im\Xi$) signals a $\lambdac$-periodic nontrivial $s_x(x)$ [$s_y(x)$] spin modulation, hence a $\lambdac$-periodic spin order.

\par

The mean-field wave functions are related to the components of the local pseudospin vector $\boldsymbol{s}(x)=\langle\hat{\boldsymbol{s}}(x)\rangle$ as,
  \begin{align}\label{eq_s}
    s_{x}(x)&=\sqrt{n_{\uparrow}(x)n_{\downarrow}(x)}\cos\Delta\phi,\nonumber\\
    s_{y}(x)&=\sqrt{n_{\uparrow}(x)n_{\downarrow}(x)}\sin\Delta\phi,\nonumber\\
    s_{z}(x)&=\frac{1}{2}[n_{\uparrow}(x)-n_{\downarrow}(x)].
  \end{align}
The normalized spin texture $\tilde{\mathbf{s}}(x)\coloneq\mathbf{s}(x)/\|\mathbf{s}(x)\|$, with $\|\mathbf{s}(x)\|=\sqrt{s_x^2(x)+s_y^2(x)+s_z^2(x)}$, for specific points in the phase diagram in Fig.~\ref{phase_diagram}(a) are shown in Figs.~\ref{spin_textures}(a)--(d).

\par

From Fig.~\ref{phase_diagram} we can, depending on the Rabi frequencies of the pump lasers, identify four distinct phases. Below threshold, i.e., for the empty cavity mode, the system is in a ferromagnetic (FM) spin state, while above threshold three different types of antiferromagnetic (AFM) ordering emerge. The two distinct areas with finite field amplitudes but opposite global magnetization $m$ [Eq.~\eqref{eq_magnetization}] correspond to the YZ-AFM and Y/XY-AFM phases. A smooth crossover between the Y-AFM and XY-AFM orderings reveals by the non-zero imaginary part of the spin order parameter $\Im\Xi$. The non-trivial phases with a finite cavity field emerge due to the polarization gradient along the cavity axis of the total electric field originating from interference of cavity and pump field.

\begin{figure*}
  \centering
  \includegraphics[width=0.95\textwidth]{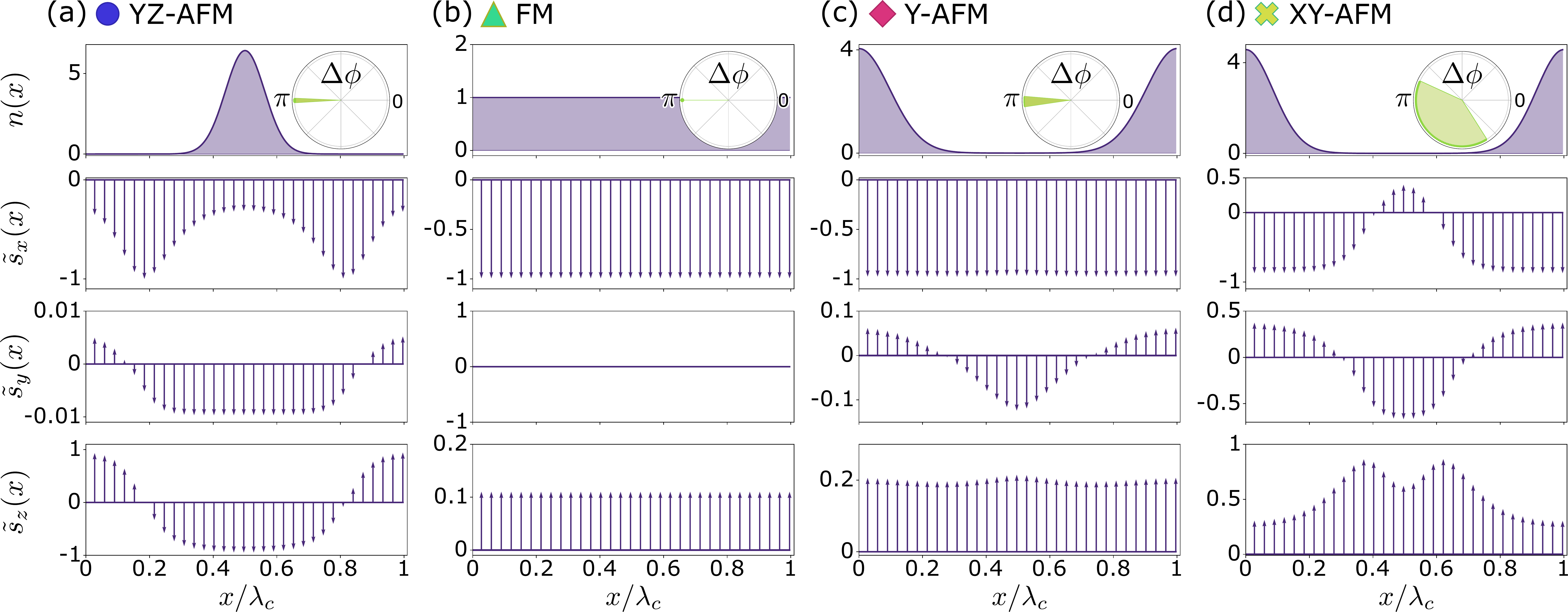}
  \caption{Total local densities $n(x)$ with insets showing the relative phase $\Delta\phi$ (first row) and textures of the normalized spin $\tilde{\mathbf{s}}(x)=\mathbf{s}(x)/\|\mathbf{s}(x)\|$ within one unit cell $\left[0,x/\lambdac\right]$ for the points corresponding to colored markers in Fig.~\ref{phase_diagram}(a):
  $(\Omega_2,\Omega_1)=(15,275)\omegar$ (YZ-AFM) (a), $(150,150)\omegar$ (FM) (b), $(250,50)\omegar$ (Y-AFM) (c), and $(275,10)\omegar$ (XY-AFM) (d), respectively. The other parameters are the same as in Fig.~\ref{phase_diagram}.}
  \label{spin_textures}
\end{figure*}

\subsection{YZ-AFM phase}

This phase appears in the left upper corner in the phase diagram in Fig.~\ref{phase_diagram}(a) where $\Omega_{1}>\Omega_{2}$. Hence, the energy-functional density $\mathcal{E}(x)$, Eq.~\eqref{energy-functional}, is dominated by the interference term $\propto\eta|\alpha|\cos\phi_{\alpha}\cos\left(\kc x\right)$ that pushes the atoms towards $x/\lambdac=\ell+1/2$ (with $\ell\in\mathbb{Z}$), see Fig.~\ref{spin_textures}(a). Such configuration is characterized by a negative value of the density order parameter $\Theta$ [Eq.~\eqref{eq_theta}], as shown in Fig.~\ref{phase_diagram}(b). This phase is very closely related to the well-known density self-ordering for two-level (i.e., effectively single-component) atoms~\cite{domokos2002collective}.

\par

The position-independent Raman coupling term $\propto\Omegap\cos\Delta\phi$ favors the relative phase of the condensate to be $\Delta\phi=\pi$. The small influence of the position-dependent Raman coupling term $\propto\Omegac|\alpha|\cos\left(\phi_{\alpha}+\Delta\phi\right)\cos\left(\kc x\right)$ slightly ``shakes'' the relative phase around $\Delta\phi=\pi$ [see the inset in Fig.~\ref{spin_textures}(a)]. This behavior is consistent with the spin structure depicted in Fig.~\ref{spin_textures}(a): Whilst the $x$ component of the spin is large and always negative, its $y$ component is tiny and changes its sign over a period, as expected for $\Delta\phi\approx\pi$ [see Eq.~\eqref{eq_s}]. Since the $y$ and $z$ components depict antiferromagnetic ordering in the deep $\lambda/2$-lattice limit we denote this phase YZ-AFM.

\par

Note that the above considerations hold for $\Re\alpha>0$; for negative $\Re\alpha$ the energy density $\mathcal{E}(x)$ favors the atoms to be located at $x/\lambdac=\ell$ (with $\ell\in\mathbb{Z}$). These two possibilities thus reflect the spontaneously-broken $\mathbb{Z}_2$ symmetry of the Hamiltonian~\eqref{eq_H}. In what follows, we likewise always consider the case $\Re\alpha>0$.

\par

One can equivalently describe the resultant magnetic order via the spin Hamiltonian $ \hat{H}_{\mathrm{spin}}$, Eq.~\eqref{H_spin}, in a heuristic manner.
The long-range cavity-mediated interactions~\eqref{coup_coeff} contained in the spin Hamiltonian~\eqref{H_spin} are mainly determined by the density-density interaction $V\left(x,x'\right)$, the $z$-component of the Heisenberg interaction $J_{z}\left(x,x'\right)$, and the interactions $W_{z}\left(x,x'\right)$ of the $z$ component of the spin with the density. The periodically modulated  coupling coefficients $\propto\cos\left(\kc x\right)\cos\left(\kc x'\right)$ induce $\lambda_\mathrm{c}$-spatial modulations in the spin components and the total density, see Fig.~\ref{spin_textures}(a), which minimize the corresponding interaction. The effective homogeneous magnetic field $\mathbf{B}$ [Eq.~\eqref{eq_B}], specially its $x$ component, also plays an important role in this case. The main influence on the $x$ component of the spin stems from the $x$ component of the effective homogeneous magnetic field $B_x$. Since $B_x$ is always positive, it is favorable that ${s}_{x}$ remains negative. The $z$ component of the magnetic field $B_z$ competes with the $z$ component of the Heisenberg interaction $J_{z}\left(x,x'\right)$ and the density-spin interactions $W_{z}\left(x,x'\right)$. The negative $B_z$ tries to align $s_z$ completely in the positive direction, resulting in the smeared $\cos(\kc x)$-periodic $s_z$ texture illustrated in Fig.~\ref{spin_textures}(a).

\subsection{FM phase}

This phase is characterized by the absence of photons in the cavity and thus appears below threshold. Hence no optical potential can build up in the cavity and the BEC stays homogeneous ($\Theta=0$). The relative phase $\Delta\phi$ of the two components is strictly locked to $\pi$, as can be seen from the inset of Fig.~\ref{spin_textures}(b) fixing the spin direction. The latter results in $\Re\Xi = \Im\Xi = 0$ and vanishing $s_{y}$. The two orthogonal spin components follow the effective external magnetic field $\mathbf{B}$, Eq.~\eqref{eq_B}, resulting in ferromagnetic order.

\subsection{Y-AFM phase}

This phase is characterized by a buildup of a coherent field in the cavity mode via Raman gain from the upper spin level. The atomic density and spin modulations are then mainly governed by the Raman coupling term [third line in Eq.~\eqref{energy-functional}] that for $\Re\alpha>0$ forces the atoms towards the even sites $x/\lambdac=\ell$ (with $\ell\in\mathbb{Z}$), see Fig.~\ref{spin_textures}(c). This is confirmed by the positive value of $\Theta$ in Fig.~\ref{phase_diagram}(b). The relative phase $\Delta\phi$ of the two condensate wave functions is again fixed around $\pi$, see inset in Fig.~\ref{spin_textures}(c), due to the competition between the position-independent Raman coupling term $\propto\Omegap\cos\Delta\phi$ and the position-dependent Raman coupling term $\propto\Omegac|\alpha|\cos\left(\phi_{\alpha}+\Delta\phi\right)\cos\left(\kc x\right)$, as in the YZ-AFM phase. However, the shaking of the relative condensate phase around $\pi$ is larger in this case, as the position-dependent Raman coupling term is bigger, leading to a more pronounced modulation in $s_y$ compared to the YZ-AFM phase.

\par

From the point of view of the spin Hamiltonian~\eqref{eq_HJVW_Hxz}, this phase is still dominated by the effective magnetic field, $|B_{x}|/\text{max}(|J_{\bot}|) \approx |B_{z}|/\text{max}(|J_{xz}|)\approx 1.7$, such that the spin orientation is mainly determined by $\mathbf{B}$. As the $y$ component of the effective magnetic field $\mathbf{B}$ is zero, $B_y=0$, $s_{y}$ is only modulated owing to $J_{\bot}\left(x,x'\right)$.

\subsection{XY-AFM phase}

Finally, we identify a fourth phase in the regime when the pump field on the cavity transition is very weak, but the Raman coupling is still strong. As for the Y-AFM phase, the atoms are mostly affected by the Raman coupling term [third line in Eq.~\eqref{energy-functional}], such that the density order parameter $\Theta$ stays positive. The Rabi frequency $\Omega_{1}$ is, however, much smaller than in the former phase such that the position-dependent term $\propto\Omegac|\alpha|\cos\left(\phi_{\alpha}+\Delta\phi\right)\cos\left(\kc x\right)$ outweighs the constant term $\propto\Omegap\cos\Delta\phi$. The minimization of the energy then results in a position-dependent relative phase $\Delta\phi(x)$, as depicted in the inset in Fig.~\ref{spin_textures}(d).

\par

For the parameters in Fig.~\ref{phase_diagram}, the pronounced spatial inhomogeneous relative phase of the two condensate wave functions  manifests itself in three ways: (i)~The $x$ ($y$) component of the spin, which is proportional to $\cos\Delta\phi(x)$ [$\sin\Delta\phi(x)$], changes its sign and exhibits strong modulations, as shown in Fig.~\ref{spin_textures}(d). (ii)~The $\lambda_\mathrm{c}$-periodic, strongly modulated $s_y$, in turn, results in non-zero $\Im\Xi$ in this phase, as depicted in Fig.~\ref{phase_diagram}(e). (iii)~The change of the sign of $\cos\left(\phi_{\alpha}+\Delta\phi\right)$ affects the atomic distribution via the Raman coupling term $\propto\sqrt{n_{\uparrow}(x)n_{\downarrow}(x)}|\alpha|\cos\left(\phi_{\alpha}+\Delta\phi\right)\cos\left(\kc x\right)$ and the energy minimization causes a local minimum in $s_z\propto n_{\uparrow}(x)-n_{\downarrow}(x)$, as shown in Fig.~\ref{spin_textures}(d). This phase is characterized by antiferromagnetic order in the spin's $x$ and $y$ components.

\par

In terms of the spin model~\eqref{H_spin}, $J_{\bot}\left(x,x'\right)$ manages to overcome the influence of the $x$ component of the effective magnetic field $\mathbf{B}$. The spatial modulation of the coupling parameter $J_{\bot}\left(x,x'\right)$ then results in the spatial modulation of $s_{x}(x)$ and $s_{y}(x)$ observed in Fig.~\ref{spin_textures}(d), except the fact that these two components are out of phase due to the presence of $B_x$. Note that both spatial dependencies of $s_{x}(x)$ and $s_{y}(x)$ minimize the corresponding interactions. The behavior of $s_{z}(x)$ is governed by the interplay of the $z$ component of the magnetic field and the cross-couplings interaction $J_{xz}\left(x,x'\right)$. As $B_z$ is always negative, $s_{z}(x)$ chooses a positive orientation. At the same time, the $\lambda_\mathrm{c}$-spatial modulations in $s_{z}(x)$ induced by $J_{xz}\left(x,x'\right)$ cause positive values of the $z$ spin component around the edges of a unit cell and negative ones close to the middle. This interplay manifests itself in the local minimum of $s_{z}(x)$ in the center of the unit cell.

\section{Conclusions}

We theoretically studied combined spin and density self-ordering of a spinor BEC inside an optical cavity, transversally illuminated by two orthogonally-polarized pump lasers in a restricted 1D geometry. We found that the long-range cavity-induced interactions among the atoms allow to engineer a broad range of density-density, spin-spin, and density-spin interactions, manifesting a rich phase diagram with different types of magnetic ordering. All magnetic phases and quantum phase transitions between them can be monitored by the cavity field leakage and atomic populations of the ground states. We have shown that despite the relative simplicity of our model, it opens an alternative way to simulate an anisotropic $t$-$J$-$V$-$W$ model together with cross-couplings between spin components. Interestingly, besides the conventional interactions between density and $z$ component of the spin, our model additionally contains interactions between the density and the spin's $x$ component. Moreover, we have investigated the various phases in terms of the energy-functional density, whose minimization dictates the spatial variation of the relative condensate phase.

\par

As a possible generalization of our scheme, we notice that in the 2D case, besides the emergence of topological spin textures, as spin spiral behavior, a more precise control of the coupling coefficients could be implemented by changing the spatial profiles of the pump fields along the $y$ axis~\cite{mivehvar2019cavity}. Furthermore, the range of the cavity-mediated density-density, spin-spin, and density-spin interactions can be tuned by exploiting a multi-mode cavity, therefore implementing tunable finite-range interactions among the atoms~\cite{vaidya2018tunable}. As a consequence of these finite-range density-density, spin-spin, and density-spin interactions, beyond mean-field phases including Luttinger and Haldane liquids emerge in the system, which will be presented elsewhere.

\par

Additional physics may also arise by including two-body contact interactions, neglected in this work. The influence of such interactions on atomic self-organization was explored in Ref.~\cite{deng2014bose}.

\section*{Acknowledgments}

We would like to thank Stefan Ostermann, Elvia Colella, and Viacheslav Kuzmin for helpful discussions. F.\,M.\ is also grateful to Luca Barbiero for fruitful discussions. This work was performed in the framework of the European Training Network ColOpt, which is funded by the European Union (EU) Horizon 2020 programme under the Marie Sk{\l}odowska-Curie action, grant agreement 721465. N.\,M.\ is supported by the Russian Foundation for Basic Research (RFBR) under the projects 19-02-00204-a and 18-02-00648-a. W.\,N.\ acknowledges support from an ESQ fellowship of the Austrian Academy of Sciences (\"OAW).  F.\,M.\ is supported by the Lise-Meitner Fellowship M2438-NBL and the international FWF-ANR Grant, No.\ I3964-N27.

\appendix

\section{Derivation of the many-body Hamiltonian}\label{A}

Within the dipole and rotating-wave approximations the single-particle Hamiltonian for the system depicted in Fig.~\ref{scheme} reads
\begin{multline}
\hat{H}_1=\frac{\hat{p}^{2}}{2m}+\sum_{\tau=\left\{ \uparrow,e\right\} }\hbar\omega_{\tau}\hat{\sigma}_{\tau\tau}+\hbar\omegac\hat{a}^{\dagger}\hat{a}\\
+\hbar\left(\Omega_{1}\hat{\sigma}_{\downarrow e}e^{i\omega_{1}t}+\mathcal{G}(\hat{x})\hat{a}^{\dagger}\sigma_{\downarrow e}+\Omega_{2}\hat{\sigma}_{\uparrow e}e^{i\omega_{2}t}+\mathrm{H.c.}\right),\label{AppA_Hinit}
\end{multline}
where $m$ is the atomic mass, $\hat{p}$ the center-of-mass atomic momentum operator along the cavity axis $x$, $\hat{\sigma}_{\tau\tau'}=|\tau\rangle\langle\tau'|$ the atomic transition operators, and $\hat{a}^{\dagger}$ the creation operator of a cavity photon. Without loss of generality we have assumed that $\left\{ \Omega_{1},\Omega_{2},\mathcal{G}_{0}\right\} \in\mathbb{R}$.

\par

The unitary transformation
\begin{equation}
  U\left(t\right)=\exp\left\{ i\left[\omega_{1}\hat{a}^{\dagger}\hat{a}+\left(\omega_{1}-\omega_{2}\right)\hat{\sigma}_{\uparrow\uparrow}+\omega_{1}\hat{\sigma}_{ee}\right]t\right\}\label{AppA_RF}
\end{equation}
transforms the Hamiltonian~\eqref{AppA_Hinit} according to $\hat{\tilde{H}}_1=U H_1 U^{\dagger}+i\hbar\left(\partial_{t}U\right)U^{\dagger}$, yielding the time-independent Hamiltonian
\begin{multline}
  \hat{\tilde{H}}_1=\frac{\hat{p}^{2}}{2m}-\hbar\Deltaa\hat{\sigma}_{ee}+\hbar\delta\hat{\sigma}_{\uparrow\uparrow}-\hbar\Deltac\hat{a}^{\dagger}\hat{a}\\
+\hbar\left(\Omega_{1}\hat{\sigma}_{\downarrow e}+\mathcal{G}(\hat{x})\hat{a}^{\dagger}\hat{\sigma}_{\downarrow e}+\Omega_{2}\hat{\sigma}_{\uparrow e}+\mathrm{H.c.}\right),\label{AppA_time_ind_H}
\end{multline}
where we have defined the detunings $\Deltaa\coloneq\omega_{1}-\omega_{e}$, $\Deltac\coloneq\omega_{1}-\omegac$ and $\delta\coloneq\omega_{\uparrow}-\left(\omega_{1}-\omega_{2}\right)$.

\par

In the large atom--pump detuning limit $|\Omega_{1,2}/\Deltaa|\ll1$ and $|\mathcal{G}_{0}/\Deltaa|\ll1$, the excited state can be adiabatically eliminated, leading to the effective pseudospin Hamiltonian
\begin{multline}\label{eq_app_Htilde1}
  \tilde{\hat{H}}_1=\frac{\hat{p}^{2}}{2m}-\hbar\Deltac\hat{a}^{\dagger}\hat{a}+\hbar\tilde{\delta}\hat{\sigma}_{\uparrow\uparrow}\\
  +\hbar\left[U_{0}\hat{a}^{\dagger}\hat{a}\cos^{2}\left(\kc \hat{x}\right)+\eta\left(\hat{a}+\hat{a}^{\dagger}\right)\cos\left(\kc \hat{x}\right)\right]\hat{\sigma}_{\downarrow\downarrow}\\
  +\hbar\Omegac\cos\left(\kc \hat{x}\right)\left(\hat{a}\hat{\sigma}_{\uparrow\downarrow}+\hat{a}^{\dagger}\hat{\sigma}_{\downarrow\uparrow}\right)+\hbar\Omegap\left(\hat{\sigma}_{\uparrow\downarrow}+\hat{\sigma}_{\downarrow\uparrow}\right).
\end{multline}
Here we have defined $\tilde{\delta}\coloneq\omega_{\uparrow}-\left(\omega_{1}-\omega_{2}\right)+\Omega_{2}^{2}/\Deltaa-\Omega_{1}^{2}/\Deltaa$, $U_{0}\coloneq\mathcal{G}_{0}^{2}/\Deltaa$, $\eta\coloneq\mathcal{G}_{0}\Omega_{1}/\Deltaa$, $\Omegac\coloneq\mathcal{G}_{0}\Omega_{2}/\Deltaa$ and $\Omegap\coloneq\Omega_{1}\Omega_{2}/\Deltaa$.

\par

Equation~\eqref{eq_H} is the many-body counterpart of the single-particle Hamiltonian~\eqref{eq_app_Htilde1}.

\par

\end{document}